\documentclass[a4paper]{article}

\usepackage{arxiv}
\usepackage[colorinlistoftodos]{todonotes}
\usepackage{pgfplots}
\usepackage{tikz} 
\usepgfplotslibrary{statistics}
\usepackage{caption}
\usepackage{tcolorbox}
\usepackage{multirow}
\usepackage{hyperref}
\usepackage{amsmath,amssymb,amsfonts,nicefrac}
\usepackage{amstext}
\usepackage{pgfplots}
\pgfplotsset{compat = 1.3}
\usepackage{multirow}
\usepackage{makecell}
\usepackage{booktabs}
\usepackage[ruled,vlined]{algorithm2e}
\usetikzlibrary{pgfplots.groupplots}

\title{MLS: A Large-Scale Multilingual Dataset for Speech Research}

\author{
Vineel Pratap, Qiantong Xu, Anuroop Sriram, Gabriel Synnaeve, Ronan Collobert \\
Facebook AI Research, Menlo Park \& Paris, USA \& France \\
\texttt{\{vineelkpratap,qiantong,anuroops,gab,locronan\}@fb.com}     
}

\begin{document}

\def\tco{\texttt{train-clean-100}}
\def\tct{\texttt{train-clean-360}}
\def\tof{\texttt{train-other-500}}
\def\devclean{\texttt{dev-clean}}
\def\devother{\texttt{dev-other}}
\def\testclean{\texttt{test-clean}}
\def\testother{\texttt{test-other}}

\maketitle
\newcommand{\gab}[1]{\todo[inline]{#1}}
\newcommand{\vineel}[1]{\todo[inline]{#1}}
\begin{abstract}
 This paper introduces Multilingual LibriSpeech (MLS) dataset, a large multilingual corpus suitable for speech research. The dataset is derived from read audiobooks from LibriVox and consists of 8 languages, including about 44.5K hours of English and a total of about 6K hours for other languages. Additionally, we provide Language Models (LM) and baseline Automatic Speech Recognition (ASR) models and  for all the languages in our dataset. We believe such a large transcribed dataset will open new avenues in ASR and Text-To-Speech (TTS) research. The dataset will be made freely available for anyone at \url{http://www.openslr.org}.

\end{abstract}

\section{Introduction}

The success of LibriSpeech~\cite{povey2015} as a standard, freely available, Automatic Speech Recognition (ASR) benchmark is undeniable in the research community. LibriSpeech is English-only, and while benchmarks for other languages are available,  there are often low-scale or scattered around different places, and rarely available under an open license. In this paper, we revisit the work which has been done with LibriSpeech but in a multi-lingual manner and at a larger scale, introducing the Multilingual LibriSpeech (MLS) dataset. MLS includes 44.5K hours of English, and a total of 6K hours spread over 7 other languages. As for LibriSpeech, MLS is a read-speech dataset, which leverages LibriVox\footnote{\url{https://librivox.org}} audiobook data, most of which being based on the
Project Gutenberg\footnote{\url{http://www.gutenberg.org}} text data. LibriVox and Project Gutenberg data are released in the public domain, which allows us to release MLS freely to everyone.

In Section 3, we detail how we created the dataset, by (i)~training some acoustic models on in-house data, (ii)~generating pseudo-labels with these models, and (iii)~retrieving the original transcript by matching pseudo-labels to available book transcripts. Section~4 details the creation of train, development and test splits and limited supervision train sets. Section~5 introduces languages models we trained for each of language. These languages models are part of the MLS release. Section~6 covers some baseline ASR experiments.


\section{Related Work}

As for our work, LibriSpeech~\cite{povey2015} is derived from the LibriVox data, and is distributed under an open license. It ships with about $1000$ hours of labeled audio, obtained by leveraging alignments between textbooks and their read (audio) counterpart. In contrast to our work, it is only mono-lingual (English). A notable multi-lingual ASR dataset was built with the IARPA Babel Program~\cite{babel}. It collected data for 24 languages, mostly from conversational telephone speech. The dataset is however not released and under an open license, and focused on low-resource languages, with labeled data ranging between $25$ to $65$ hours per language. On the open license side, two important volunteer-supported multi-lingual speech gathering efforts are being conducted: (i)~VoxForge~\cite{Voxforge.org} which collected data for about 15 different languages, but remains low-scale (about $300$ hours in total). (ii)~CommonVoice~\cite{ardila2019common}, a more scalable solution, with more than 30 languages available, which keeps growing with $4500$ (validated) hours currently available. Other notatable multi-lingual datasets distributed under an open license are the M-AILABS~\cite{ailabs2019} and the CMU Wilderness~\cite{black2019} datasets. M-AILABS is a lower-scale version of our work, with 9 languages collected from LibriVox, for a total of about $1000$ hours available. The CMU Wilderness collects readings from the New Testament, with 700 different languages available.


\section{Data processing pipeline}
This section describes the major steps involved in preparing the MLS dataset. 

\begin{table*}
\centering
 \caption{LibriVox audiobooks statistics for the top 15 languages (* - audiobooks with mix of multiple languages)} \label{tab:lv_data}
  \begin{tabular}{lrrr}
    \toprule
    \textbf{Language}      & \textbf{Hours}  & \textbf{Books}     & \textbf{Speakers}                \\
    \midrule
	English	& 71,506.79&12421&		4214 \\
	German	& 3,287.48& 593&		244 \\
	Dutch&		2,253.68& 206&	91 \\
	Spanish& 1,438.41&	285&		120 \\
	French&		1,333.35& 224&	114 \\
	Multilingual*& 516.82&	130	&	19 \\
	Portuguese& 	284.59	&	68&31 \\
	Italian	& 	279.43& 61&	28 \\
	Russian& 172.34&	 	44&	29 \\
	Latin& 138.93&		20&	16 \\
	Polish&		137.00& 25&	16 \\
	Church Slavonic&	136.42&		8&2 \\
	Hebrew	&125.72& 23	&	13 \\
	Japanese&97.67&	38&		24 \\
	Ancient Greek&69.77&	43&		8 \\
    \bottomrule
  \end{tabular}
\end{table*}

\subsection{Downloading audiobooks}
Table \ref{tab:lv_data} shows the LibriVox audiobooks data available for each language measured using LibriVox APIs~\footnote{\url{https://librivox.org/api/info}}. While English is the most dominant language, there is a significant amount of audio hours present in languages other than English, making this a valuable source for multilingual dataset preparation. 

Based on the number of audiobook hours and the availability of the corresponding text sources of the audiobooks, we have selected English, German, Dutch, Spanish, French, Portuguese, Italian, Polish for the MLS dataset preparation. For downloading the LibriVox audiobooks in these languages, we have used data preparation tools available at Libri-Light~\footnote{\url{https://github.com/facebookresearch/libri-light}} open source library. All the audio data is downsampled from 48kHz to 16kHz for further processing. 

\subsection{Audio segmentation}
Audio files downloaded from LibriVox have a huge variance in duration - from few minutes to hours.  Since acoustic model training is usually done on shorter utterances, we segment the audio files into 10-20 second segments. For this, we have used trained acoustic models in each of the languages using Time-Depth Separable (TDS) Convolutions\cite{Hannun2019SequencetoSequenceSR} with Auto-Segmentation Criterion~\cite{collobert2016wav2letter} (ASG) loss. We chose ASG criterion over Connectionist Temporal Classification~\cite{graves2006connectionist} (CTC) criterion since ASG doesn't exhibit delay in transcriptions compared to CTC~\cite{liptchinsky2017letterbased}. For each language, we train models on in-house datasets consisting of videos publicly shared by users. We use only audio part of the videos and the data is completely de-identified. 

\begin{figure}
\centering
   \includegraphics[width=\linewidth]{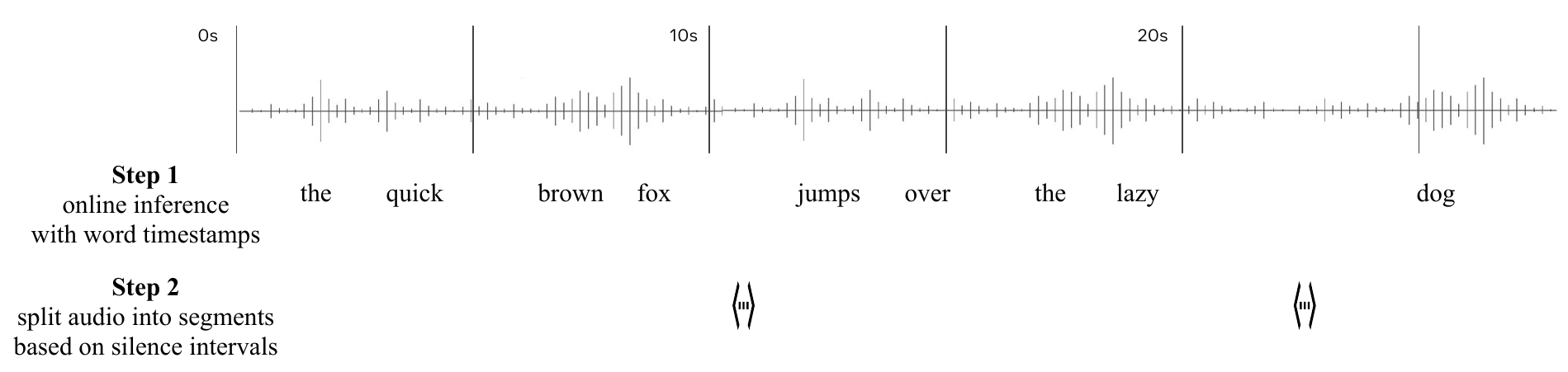}
  \caption{Audio segmentation procedure to split long audio files into shorter segments}
  \label{fig:segment}
\end{figure}

Our segmentation process is shown in Figure \ref{fig:segment} and consists of two steps. First, we run inference on the audio and generate viterbi token sequence along with their timestamps. Since the audio files can be very long, we have used wav2letter@anywhere framework~\cite{pratap2020scaling} to perform the inference in a streaming fashion. Second, we select the longest silence duration within 10 sec to 20 sec range from the start of an audio and split the audio at the mid point of the silence chunk to create a segment. If no silence frames are found between 10 sec to 20 sec from the starting point, we split the audio at 20 sec mark. Once we generate a segment, we consider the end point of previous segment as the current starting point and repeat the process again till we reach the end of audio.  This process guarantees that all the segments are between 10sec and 20sec. A minimum segment duration of 10 sec is kept so that the segments have sufficient number of words spoken which helps with better transcript retrieval (described in Section ~\ref{trans_ret}). 

\subsection{Pseudo label generation}
\label{sec:pl}
We generate pseudo labels for the segmented audio samples by performing a beam-search decoding with a 4-gram language model on the same acoustic models used for audio segmentation. The language models are trained on the data used for training the acoustic models. For English, however, we use pre-trained model from~\cite{synnaeve2019endtoend} which uses TDS encoder and CTC loss on LibriSpeech~\cite{panayotov2015librispeech} and pseudo labels from LibriVox . 

\subsection{Downloading text sources for audiobook data}
To generate the labels for the audio segments derived from audiobooks, we would need to have the original textbook from which the speaker read the audiobook. For English, we found that $\approx$60K hours of audiobooks is read from four major website domains - \url{gutenberg.org}, \url{archive.org}, \url{ccel.org} and \url{hathitrust.org}. We have written parsers to automatically extract the text for each of these domains and downloaded the text sources for all the audiobooks in English. 

For other languages, however, we found it more challenging to automatically extract text for the audiobooks because: 1. the diversity of domains is large making it difficult to write parsers for each and every domain 2. some of the links were invalid or redirected to an incorrect page. So, we have also incorporated some manual approaches in our process to cover the audiobook text sources as much as possible. Depending on the language and text source domain, we copied the text data directly from the  browser or extracted text from .pdf/.epub books using pdftotext\footnote{\url{https://pypi.org/project/pdftotext/}} tool, or wrote HTML parsers to retrieve text data for popular domains in a language. 

For the audiobooks with invalid links for text sources, we have also manually searched online to find alternate sources where the book is available. For example, all the text sources from the \url{spiegel.de} domain, which accounts of 1/3rd of German audio data, were being redirected to an invalid page. However, we were able to find the alternate text sources from websites like \url{projekt-gutenberg.de}, \url{zeno.org} for most of these unavailable books from \url{spiegel.de}. 

\subsubsection{Text Normalization}
\label{tnorm}
Given that the text data is taken from many different online sources, we perform careful text normalization to make the data easy to work with for later stages of the data processing pipeline. For normalizing the text, we first perform NFKC normalization and remove all the unwanted characters like punctuations, subscript/superscripts, emojis, escape symbols etc.  We also remove the hyphens used for ``end-of-line hyphenation"~\footnote{\url{https://practicaltypography.com/hyphenation.html}} and join the parts of words into a single word. We then prepare a list of valid unicode characters based on the language’s orthography and filter characters outside this range. 

\subsection{Transcript retrieval}
\label{trans_ret} 

The transcript retrieval process involves finding the true target label for the audio segments from the source text of audio. Our procedure closely follows the method described in~\cite{manohar2017jhu} with few modifications. 

We first split the source text into multiple overlapping documents of 1250 words each and striding by 1000 words. We retrieve the documents which best matches with the pseudo label for the audio segments using term-frequency inverse document-frequency (TF-IDF) similarity score on bigrams. We then perform a Smith-Waterman alignment~\cite{SMITH1981195} to find the best matching sub-sequence of words. We have used a matching score of 2 and substitution, insertion, deletion score of -1 for the alignment algorithm. 

After the above alignment procedure, we generate a candidate target label for each audio segment, which corresponds to the best match of the pseudo label in the source text of audiobook. We filter out all the candidate transcripts generated from the matching algorithm above, if the WER between the candidate transcript and pseudo label generated is \textgreater 40\%.


\subsubsection{Post processing of numbers, hyphens, apostrophe }

\begin{figure}
\begin{minipage}{.48\textwidth}
  \centering
    \caption{Example implementation of number replacement procedure. Numbers from book text are replaced with the aligned word from pseudo label.}
     \includegraphics[width=0.95\linewidth]{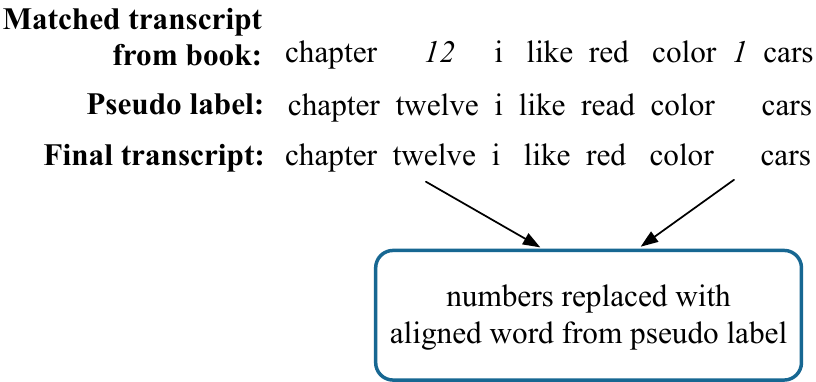}
  \label{fig:numbers}
\end{minipage}%
\hfill
\begin{minipage}{.48\textwidth}
  \centering
    \caption{Heuristics used for dealing with hyphens and apostrophes. Words marked in bold are replaced with the words present below them.}
     \includegraphics[width=0.95\linewidth]{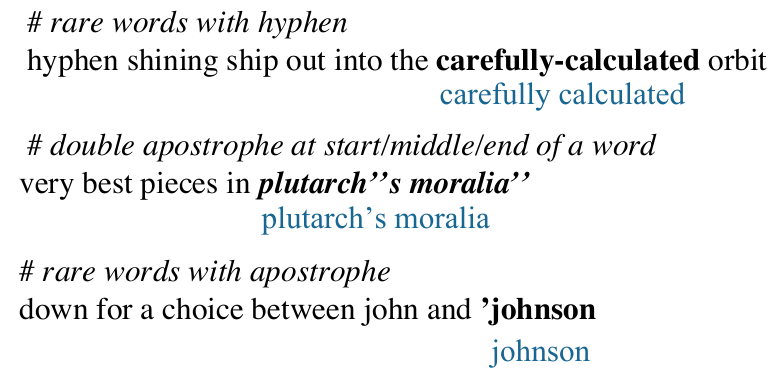}
  \label{fig:hyphens}
\end{minipage}
\end{figure}

Numbers in the text present a unique challenge in the preparation of the dataset. While a library like num2words\footnote{\url{https://pypi.org/project/num2words/}} can be used to convert numbers to words and supports multiple languages, we found that using such tools won't solve the problem in a generic way. For example, a number ``401" could be spelled as ``four hundred and one"  or  ``four-o-one" depending on the context and it is very hard to come up with automated rules to pick one over the other. Further, some of the numbers present in the text could be page numbers and are not read by the speaker. To solve the issue, we align the matched text from the book with the pseudo label and replace the numbers in the matched text with the corresponding aligned words from the pseudo label. While this solution is not perfect as the pseudo label may be not be always have the correct transcription of the audio, we found this to be a reasonably good solution to solve the issue. Figure \ref{fig:numbers} shows an example implementation of this procedure. 

On the other hand, dealing with hyphens and apostrophes is challenging because we found these character to appear in the text in many places where we do not expect them to be for an audio transcript. Figure \ref{fig:hyphens} shows some examples of these issues from the books we processed and the heuristics we used to replace such word forms. To determine is a word is rare, we count of number of distinct books the word appears and check if it is less than a chosen threshold.

\section{Putting it together}
\subsection{Creating train, development and test splits}

\begin{table*}[]
\centering
\caption{Statistics of Train/Dev/Test partitions of each language. Below lists for each partition: the total duration in hours (left), number of speakers in each gender (middle) and duration per gender in dev and test sets (right). \label{tab:speaker_stat}}
\begin{tabular}{@{}cccccccccccccc@{}}
\toprule
\multirow{3}{*}{\textbf{Language}} & \multicolumn{3}{c}{\textbf{Duration (hrs)}}                                          & \multicolumn{6}{c}{\textbf{\# Speakers} }                                                & \multicolumn{4}{c}{\textbf{\# Hours / Gender}}             \\ \cmidrule(lr){2-4} \cmidrule(lr){5-10} \cmidrule(lr){11-14}
                          & \multirow{2}{*}{train} & \multirow{2}{*}{dev} & \multirow{2}{*}{test} & \multicolumn{2}{c}{train} & \multicolumn{2}{c}{dev} & \multicolumn{2}{c}{test} & \multicolumn{2}{c}{dev} & \multicolumn{2}{c}{test} \\ \cmidrule(lr){5-6} \cmidrule(lr){7-8} \cmidrule(lr){9-10} \cmidrule(lr){11-12} \cmidrule(lr){13-14}
                          &                        &                      &                       & M           & F           & M          & F          & M           & F          & M          & F          & M           & F          \\
\midrule                          
English    & 44,659.74 & 15.75 & 15.55 & 2742 & 2748 & 21 & 21 & 21 & 21 & 7.76 & 7.99 & 7.62 & 7.93 \\
German     & 1,966.51  & 14.28 & 14.29 & 81   & 95   & 15 & 15 & 15 & 15 & 7.06 & 7.22 & 7.00    & 7.29 \\
Dutch      & 1,554.24  & 12.76 & 12.76 & 9    & 31   & 3  & 3  & 3  & 3  & 6.44 & 6.32 & 6.72 & 6.04 \\
French     & 1,076.58  & 10.07 & 10.07 & 62   & 80   & 9  & 9  & 9  & 9  & 5.13 & 4.94 & 5.04 & 5.02 \\
Spanish    & 917.68    & 9.99  & 10    & 36   & 50   & 10 & 10 & 10 & 10 & 4.91 & 5.08 & 4.78 & 5.23 \\
Italian    & 247.38    & 5.18  & 5.27  & 22   & 43   & 5  & 5  & 5  & 5  & 2.5  & 2.68 & 2.38 & 2.90  \\
Portuguese & 160.96    & 3.64  & 3.74  & 26   & 16   & 5  & 5  & 5  & 5  & 1.84 & 1.81 & 1.83 & 1.90  \\
Polish     & 103.65    & 2.08  & 2.14  & 6    & 5    & 2  & 2  & 2  & 2  & 1.12 & 0.95 & 1.09 & 1.05 \\            
\bottomrule
\end{tabular}
\end{table*}

We have used the following principles when splitting the dataset into train, development and test sets - 1) there is no speaker overlap between the training, development and test sets, 2) speakers are balanced in gender and duration in development and test sets and 3) there are sufficient audio hours and speakers assigned into development and test sets to be able to validate ASR model performance. 

First, we select the list of all the books available for a language. We remove all the books with corrupted meta data, such as missing title or information about speakers and authors. Then, to ensure that each recording is unambiguously attributable to a single speaker, we also remove audios with multiple speakers, for example, “Dramatic Reading”, which include predominantly multi-reader audio chapters. In addition, we only keep the latest version for books sharing the same authors and title, but different versions. Only the speakers reading those valid books are considered in the transcript retrieval process from section \ref{trans_ret}.

Second, we label the gender of all the speakers with a gender classifier. The classifier is a SVM \cite{svm} with RBF kernel trained on 1172 speakers from \textit{train-clean-100} and \textit{train-clean-360} subsets of LibriSpeech . In particular, it consumes 40-dimensional log-filterbank features averaged over time as input features. The test accuracy on the 146 speakers from the joint development and test sets of LibriSpeech is 95\%. We use the same gender classifier for other languages as well. We manually checked the quality of this classifier on Dutch and Polish and its accuracy is 96\% and 94\% respectively. 

Finally, the dataset is split into training, development and test sets as following. We computed the total duration each speaker spends in reading the valid books, and order the speakers by this duration. Speakers with duration shorter than a threshold are assigned into the training set. Then, from the rest of the speakers, we select a series of speakers with the shortest duration in each gender equally. The selected speakers are then equally split into development and test sets. All the remaining speakers are assigned to training set again. Finally, after the speakers are assigned to their given partitions, we simply attach the samples recorded by them into different partitions accordingly. To avoid high speaker imbalance in development and test sets, we further truncate speakers with high duration by sampling their recordings up to an upper-bound. We also make sure that each chapter of each valid book only appears in one partition. The detailed statistics can be found in Table \ref{tab:speaker_stat}. 

\begin{figure}
\centering
\begin{minipage}{.48\textwidth}
  \centering
    \caption{Speaker (fully filled), chapter (gridline filled) and book (empty) overlap of MLS and LibriSpeech (LS) partitions for English. The figure is not to scale.}
   \includegraphics[width=0.95\linewidth]{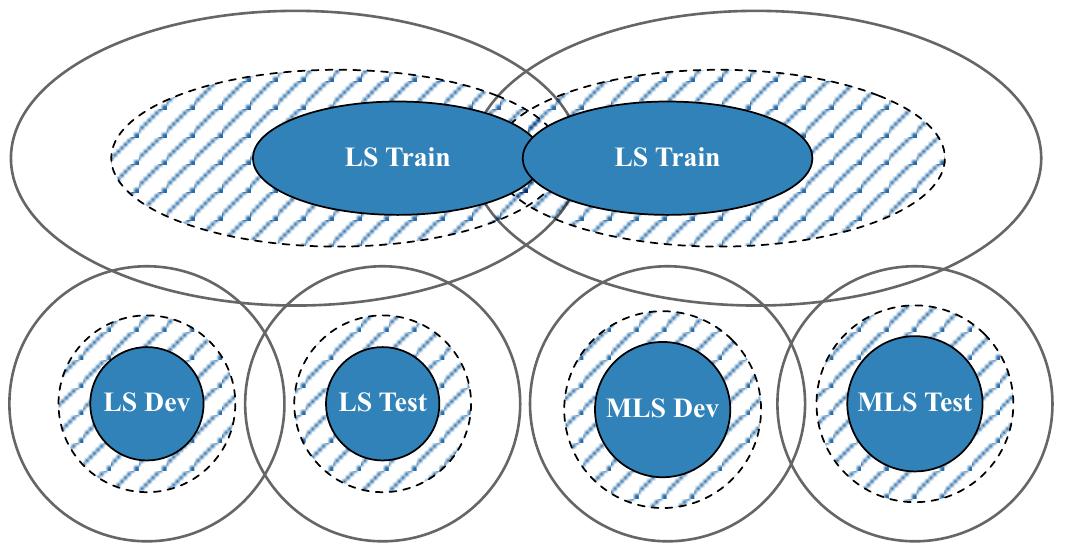}
  \label{fig:partitions}
\end{minipage}%
\hfill
\begin{minipage}{.48\textwidth}
  \centering
    \caption{Violin plots of  audio segments duration in the training data for different languages}
    \includegraphics[trim={1.1cm 0.5cm 1.8cm 0.5cm},clip, width=\linewidth]{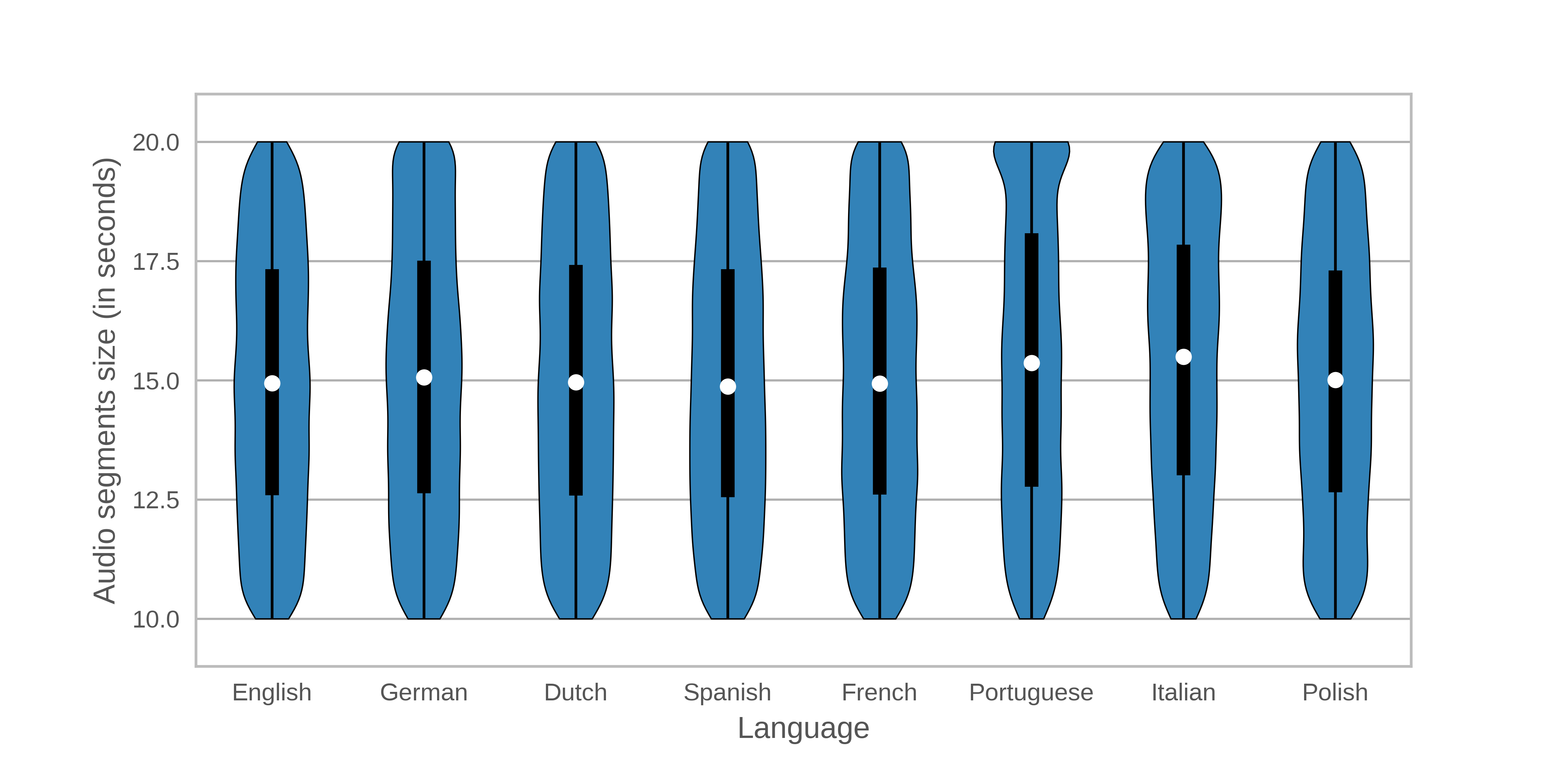}

  \label{fig:violin}
\end{minipage}
\end{figure}

For English, aside from the same procedure described above, we additionally make it exclusive to the previous LibriSpeech dataset. As shown in Figure \ref{fig:partitions}, there is speaker and book overlap between the two train sets, but the development and test sets are completely independent from one another. This makes the 2 training sets interchangeable and complementary. Note that there is book overlap across partitions, because following LS preparation, different chapters of a given book may be read by different speakers. In addition, the MLS development and test sets are chosen to be harder than LibriSpeech dev-/test-other paritions. Specifically, by comparing the pseudo labels generated as section \ref{sec:pl} with the final transcriptions, speakers with WER higher than 80\% of the ones in dev-other sets are picked into MLS dev/test sets. 

\subsection{Limited supervision train data}
We also provide limited supervision dataset for each language to create a standard benchmark for low resource training with MLS. Similar to \cite{9052942}, we provide a 10 hour set, a 1 hour set, and six 10-minute sets (the six 10-minute sets together make up the 1h set, and the 1h set is included in the 10h set). For preparing the six 10-minute sets, we first sample upto 15 male and 15 female speakers from train set for each language. From this set of speakers, we select 3 male speakers and 3 female speakers at random and sample 5 minutes of audio for each gender making sure we don't select a sample from a previous 10 min train set. For preparing the remaining 9 hour set, we consider all the speakers selected initially and sample 4.5 hours of audio per gender that is not present in the six 10-minute sets created before.

\subsection{High quality development and test sets}
Using a high quality development and test sets are critical for measuring ASR system performance and making decisions based on the test results. In order to make sure the quality of MLS development and test sets is high, we perform human evaluation of the transcripts in these subsets. The human transcribers listen to the audio file while looking the transcript and are asked to correct the transcripts if they find any mistakes. To ensure that we do not add any potential bias into the dataset from human transcribing, we ask the transcribers to only correct ``clear and obvious" errors. 

Figure \ref{fig:human_corrections} shows some of the examples of corrections using this workflow and Table \ref{tab:human_wer} shows the WER between generated transcript and human rated transcript.

\begin{table}
\centering
\begin{minipage}{.48\textwidth}
  \centering
    \captionof{figure}{Example corrections to generated transcriptions made as part of human verification workflow. GT - Generated Transcript, HRT - Human Rated Transcript}
      \begin{tcolorbox}[colback=gray!5!white,colframe=white!20!black,boxrule=0.5pt,arc=0pt, boxsep=2pt,left=2pt,right=2pt,top=2pt,bottom=2pt]
   \textit{\# Diactric correction} \\
   GT:  menesteroso para matar á los de recto proceder \\
   HRT: menesteroso para matar a los de recto proceder \\
  
   \textit{\# Missing words} \\
   GT: cabeça escandecida amor que vive\\
   HRT: cabeça escandecida amor que vive e brilhe \\
   
   \textit{\# Missing Apostrophe} \\
   GT: teseo combatter co doppi petti e degli ebrei\\
   HRT: teseo combatter co' doppi petti e degli ebrei
\end{tcolorbox}

  \label{fig:human_corrections}
\end{minipage}%
\hfill
\begin{minipage}{.48\textwidth}
\centering
\caption{WER between generated transcript (GT) and human rated transcript (HRT). GT is considered as hypothesis and HRT as reference for measuring WER.} 
\begin{tabular}{ccc}
    \toprule
        \multirow{2}{*}{\textbf{Language}} & \multicolumn{2}{c}{\textbf{Error Rate}} \\ 
 \cmidrule(lr){2-3}
        & Dev & Test \\
    \midrule
     English&	4.55 & 	4.54 \\
German & 	3.6 & 	3.9 \\
Dutch & 	7.24 & 	8.61 \\
French & 	3.33 & 	4.45 \\
Spanish & 	1.93 & 	3.07 \\
Italian & 	6.8 & 	5.37 \\
Portuguese & 	12.64 & 	12.07 \\
Polish & 	5.67 & 	5.12 \\
    \bottomrule
\end{tabular}
\label{tab:human_wer}
\end{minipage}
\end{table}

\subsection{Dataset statistics} 
Table \ref{tab:speaker_stat} shows the amount of train, valid and test data  along with the gender distribution of speakers for the 8 languages that we processed.  The training data is lower than the data we presented in Table \ref{tab:lv_data} because of 1) we may not have downloaded the text source for the audiobook, 2) the candidate label is filtered during the transcript retrieval stage or 3) filtered while creating valid/test splits. 


 Figure~\ref{fig:violin} shows the violin plots of the duration of audio segments in training data for each language. We can see that all the segments are within 10sec to 20 sec range and all the sizes are almost evenly distributed inside the range.

\section{Language models}
\label{sec:lm}
We also release language model training data and pre-built language models along with the MLS dataset. The source text for these language models is taken from Project Gutenberg\footnote{http://gutenberg.org/} books and the text is carefully normalized similar to the steps mentioned in Section \ref{tnorm}. 

We have also carefully filtered certain books to avoid any overlap with the development and test sets, following the procedure described in \cite{panayotov2015librispeech}. Specifically, any book whose title has edit distance (over words) less than $2$ with any of the titles of books in development and test sets are removed in their entirety. Then, an inverted index of all 5-grams, with stop words removed, is built for the books in the test and development sets. All candidate books are then checked against this index and any book with more than one percent of the 5-grams which appear in it is removed as well.

We have trained 3-gram and 5-gram language models (LM) for all the languages in our dataset using the KenLM toolkit~\cite{heafield2011kenlm}. We release all the language model training data and pre-built language models with the dataset. The detailed information of language model corpus for each language can be found in Table \ref{tab:lm_corpus}. The  out-of-vocabulary (OOV) rate and their perplexities (excluding OOV words) on the transcriptions in development sets are listed in Table \ref{tab:lm_ppl}. 

\begin{table*}[]
\centering
\caption{Language model corpus statistics. Number of books before and after filtering (col 2 - 3), number of words (col 4 - 5) and number of sentences (col 6). \label{tab:lm_corpus}}
\begin{tabular}{@{}cccrrr@{}}
\toprule
\textbf{Language} & \textbf{\# Books} & \textbf{\# Books (after filtering)}  & \textbf{\# Words} & \textbf{\# Words (Unique)} & \textbf{\# Sentences} \\ 
\midrule
English           & 47677            & 36866 (77\%)                 & 2.38B         & 4.12M                     & 133.26M             \\
German            & 1757             & 1323 (75\%)                 & 67.06M           & 1.24M                     & 3.81M               \\ 
Dutch             & 796              & 629 (79\%)                 & 36.26M           & 559.6K                      & 1.81M               \\
French            & 3008             & 2424 (81\%)                 & 164.1M          & 911.22K                      & 8.34M               \\
Spanish           & 622              & 479 (77\%)                 & 32.41M           & 489.94K                      & 1.61M               \\
Italian           & 814              & 640 (79\%)                 & 42.28M           & 700.63K                      & 2.14M               \\
Portuguese        & 554              & 460 (83\%)                 & 13.84M           & 383.2K                      & 1.24M               \\
Polish            & 31               & 22 (71\%)                 & 492.32K             & 67.1K                       & 28.86K             \\
\bottomrule
\end{tabular}
\end{table*}

\begin{table*}[]
\centering
\caption{The performance of the language models of each language.  Out-of-vocabulary (OOV) rates on dev sets (col 2) and perplexity on dev sets for 3-gram and 5-gram LMs (col 3). \label{tab:lm_ppl}}
\begin{tabular}{@{}cccc@{}}
\toprule
 & & \multicolumn{2}{c}{\textbf{LM Perplexity}} \\
\cmidrule(lr){3-4}
\multirow{-2}{*}{\textbf{Language}} & \multirow{-2}{*}{\textbf{OOV rate}} & 3-gram  & 5-gram \\
\midrule
English     & 0.18\% & 232.11 &   190.76            \\
German      & 1.22\% & 592.22  &   570.84            \\
Dutch       & 1.50\% & 461.22 &   439.62            \\
French      & 0.42\% & 261.65 &   238.47            \\
Spanish     & 0.92\% & 376.81       &    358.09     \\
Italian     & 1.72\% &  733.10      &     632.53    \\
Portuguese  & 1.81\% & 983.71       &    949.96   \\
Polish      & 13.39\%  &  2450.23      &   2442.12  \\
\bottomrule
\end{tabular}
\end{table*}

\section{Experiments and results} 

\subsection{Monolingual Baselines}

All the baseline models are trained using wav2letter++ \cite{pratap2018} framework. The  network architecture of acoustic model closely follows \cite{likhomanenko2020rethinking}: the encoder of our AMs is composed of a convolutional frontend (1-D convolution with kernel-width 7 and stride 3 followed by GLU activation) followed by 36 4-heads Transformer blocks\cite{vaswani2017attention}. The self-attention dimension is 768 and the feed-forward network (FFN) dimension is 3072 in each Transformer block. The output of the encoder is followed by a linear layer to the output classes. We use dropout after the convolution layer. For all Transformer layers, we use dropout on the self-attention and on the FFN, and layer drop~\cite{fan2020}, dropping entire layers at the FFN level. Dropout and layer dropout values are tuned for each model separately. 

We use the Adagrad optimizer~\cite{duchi2011adaptive} and decay learning rate by a factor of 2 each time the WER reaches a plateau on the validation sets.
The AMs take 80-channel log-mel filterbanks as input and are trained end-to-end with Connectionist Temporal Classification (CTC) loss~\cite{graves2006connectionist}. SpecAugment\cite{Park_2019} is used for data augmentation in training: there are two frequency masks, and ten time masks with maximum time mask ratio of p = 0.1; frequency and time mask parameters are tuned separately for each model; time warping is not used. The maximum frequency bands masked by one frequency mask is 30, and the maximum frames masked by the time mask is 30, too. We use the set of graphemes from the corresponding training set of the language as the token set. The viterbi WER on development and test for all the languages are shown in Table \ref{tab:baslines}. 

To further improve the WER, we use beam-search decoding in wav2letter++ without using a LM (ZeroLM) and using 5-gram language model mentioned in Section \ref{sec:lm}. The decoder hyper-parameters are tuned on the development set and the results are shown in Table \ref{tab:baslines}. We can see that WER improves when decoding with a 5-gram LM for all the languages except Polish. For Polish, the OOV rate is 13\% as shown in Table \ref{tab:lm_ppl} which negatively impacts the WER as we are performing LM decoding with constrained lexicon. All the pretrained models and the recipes to train the acoustic model and decoding them will be made available at \url{https://github.com/facebookresearch/wav2letter}.

\begin{table*}[]
\centering
\caption{Monolingual Baseline WER with different decoding strategies. \label{tab:baslines}}

\begin{tabular}{@{}ccccccc@{}}
\toprule
                                    & \multicolumn{2}{c}{\textbf{Viterbi}} & \multicolumn{2}{c}{\textbf{Zero LM}} & \multicolumn{2}{c}{\textbf{5-gram LM}} \\ \cmidrule(lr){2-3} \cmidrule(lr){4-5} \cmidrule(lr){6-7}

\multirow{-2}{*}{\textbf{Language}} & dev               & test             & dev                          & test  & dev                 & test                \\ 
\midrule
English                             & 6.01              & 6.99             & 5.89                         & 6.76  & 5.05                & 5.88                \\
German                              & 5.55              & 6.93             & 5.80  & 7.10   & 5.45                & 6.49                \\
Dutch                               & 17.00                & 13.18            & 16.45                        & 13.09 & 14.42               & 12.02               \\
French                              & 7.79              & 6.88             & 7.43                         & 6.58  & 6.58                & 5.58                \\
Spanish                             & 5.94              & 6.90              & 5.90                          & 6.68  & 5.27                & 6.07                \\
Italian                             & 14.55             & 12.35            & 14.01                        & 11.78 & 11.96               & 10.54               \\
Portuguese                          & 18.62             & 21.70             & 17.22                        & 20.52 & 16.35               & 19.49               \\
Polish                              & 19.25             & 19.40             & 18.73                        & 21.66 & 17.64               & 20.39              \\
\bottomrule
\end{tabular}
\end{table*}

\subsection{Comparison with LibriSpeech}

\begin{table*}[]
\centering
\caption{Comparing model performance on Librispeech dev/test sets using LibriSpeech and MLS (different amount of training data, same model setup). \label{tab:ls_wer}}
\begin{tabular}{@{}cccccccc@{}}
\toprule
\multirow{2}{*}{\textbf{Training Dataset}} &
\multirow{2}{*}{\textbf{Training Hours}}&
\multicolumn{2}{c}{\textbf{Viterbi}} & \multicolumn{2}{c}{\textbf{4-gram LM}} & \multicolumn{2}{c}{\textbf{Transformer LM}} \\ \cmidrule(lr){3-4} \cmidrule(lr){5-6} \cmidrule(lr){7-8}
                                      &  & clean             & other            & clean                    & other                   & clean                     & other                     \\
\midrule
LibriSpech  \cite{likhomanenko2020rethinking}      &     960                   & 2.8 / 2.8           & 7.1 / 7.1          & 2.0 / 2.5                  & 5.3 / 5.6                 & 1.5 / 2.1                   & 4.3 / 4.7                   \\
MLS                   &  44.5K         & 2.94 / 3.09       & 5.39 / 5.50       & 2.06 /2.33                & 4.01 / 4.38               & 1.83 / 2.11                 & 3.54 / 3.97                \\
\bottomrule
\end{tabular}
\end{table*}

Table \ref{tab:ls_wer} compares the WER performance on LibriSpeech dev and test sets using the same model architecture. We use the 4-gram LM provided with LibriSpeech for decoding both models and the transformer LM used for rescoring is trained on the LM corpus and training data provided with LibriSpeech. It can be seen that using MLS dataset for English and thereby using large amount of supervised training data improves the performance on LibriSpeech dev and test sets. 




\section{Conclusions}

We have presented the Multilingual LibriSpeech dataset, a large scale multilingual speech dataset with 36.5K hours of training data spread over 8 languages. We believe this dataset will promote open research in large-scale training of ASR systems and in multilingual ASR. This dataset can also be used for Text-to-Speech (TTS) research by extending the LibriTTS~\cite{48008} dataset, and by creating a larger and multilingual version for TTS Research. 
\section{Acknowledgements}
We would like to thank Steven Garan for help in data preparation and text normalization and Mark Chou for helping with setting up the workflow for transcription verification. 

\newpage

\bibliographystyle{IEEEtran}
\bibliography{mybib}

\end{document}